\input{aipcheck}

\documentclass[
    ,final            
  ]
  {aipproc}

\layoutstyle{6x9}

\def\met{\mbox{$\rm {\hbox{E\kern-0.4em\lower-.1ex\hbox{/}}}_T$}}
\def\pet{\mbox{$\rm {\hbox{P\kern-0.4em\lower-.1ex\hbox{/}}}_T$}}
\def\metx{\mbox{$\rm {\hbox{E\kern-0.4em\lower-.1ex\hbox{/}}}_x$}}
\def\mety{\mbox{$\rm {\hbox{E\kern-0.4em\lower-.1ex\hbox{/}}}_y$}}
\begin{document}

\title{Searches for Beyond the Standard Model Higgs Bosons in $\mathbf{p\bar{p}}$ collisions at
$\mathbf{\sqrt{s}}$=1.96 TeV}


\author{Prolay Kumar Mal\\
(On behalf of D\O\  Collaboration)}{
  address={University of Arizona\\
Tucson, Arizona 85721, USA}
}

\begin{abstract}
The recent results on various Beyond the Standard Model (BSM)
Higgs boson searches performed by the D\O\  experiment at the
Tevatron are presented here. In particular, the Higgs bosons in
supersymmetric models and fermiophobic scenario have been
investigated. No significant excess over the Standard Model (SM)
expectations have been observed and accordingly limits have been
established on the corresponding model parameters.
\end{abstract}

\maketitle


\section{Introduction}
Besides the Standard Model Higgs Boson searches, the D\O\ 
physics program is enriched with various well motivated
BSM Higgs boson searches. The searches are primarily
focused on the Higgs bosons from three different models:
the Fermiophobic Higgs model, the
Minimal Supersymmetric Standard Model or MSSM
and the next to Minimal Supersymmetric Standard Model or
NMSSM. The analyses summarized here
are performed with 1-4.2 $\rm fb^{-1}$ of $\rm p\bar{p}$
collision data recorded with the D\O\  detector during
Run II of the Tevatron Collider at Fermilab.

\section{Fermiophobic Higgs Search}
The Fermiophobic model assumes zero couplings of the Higgs boson
to fermions while the Higgs couplings to the gauge bosons remain
the same as in the SM. In such models, Higgs boson production via
gluon fusion is absent and the Higgs bosons are mainly produced
in association with a $\rm W^\pm$ or a $\rm Z^0$ boson (VH) and via
vector boson fusion (VBF). D\O\  has performed the searches for such
Fermiophobic Higgs bosons ($\rm h_f$) in two different channels:
$\rm h_f\rightarrow\gamma\gamma$ and $\rm W^\pm h_f(\rightarrow W^+W^-)$
with 4.2 $\rm fb^{-1}$ and 3.6 $\rm fb^{-1}$ datasets respectively.

\begin{figure}
\begin{tabular}{cc}
\includegraphics[height=.15\textheight]{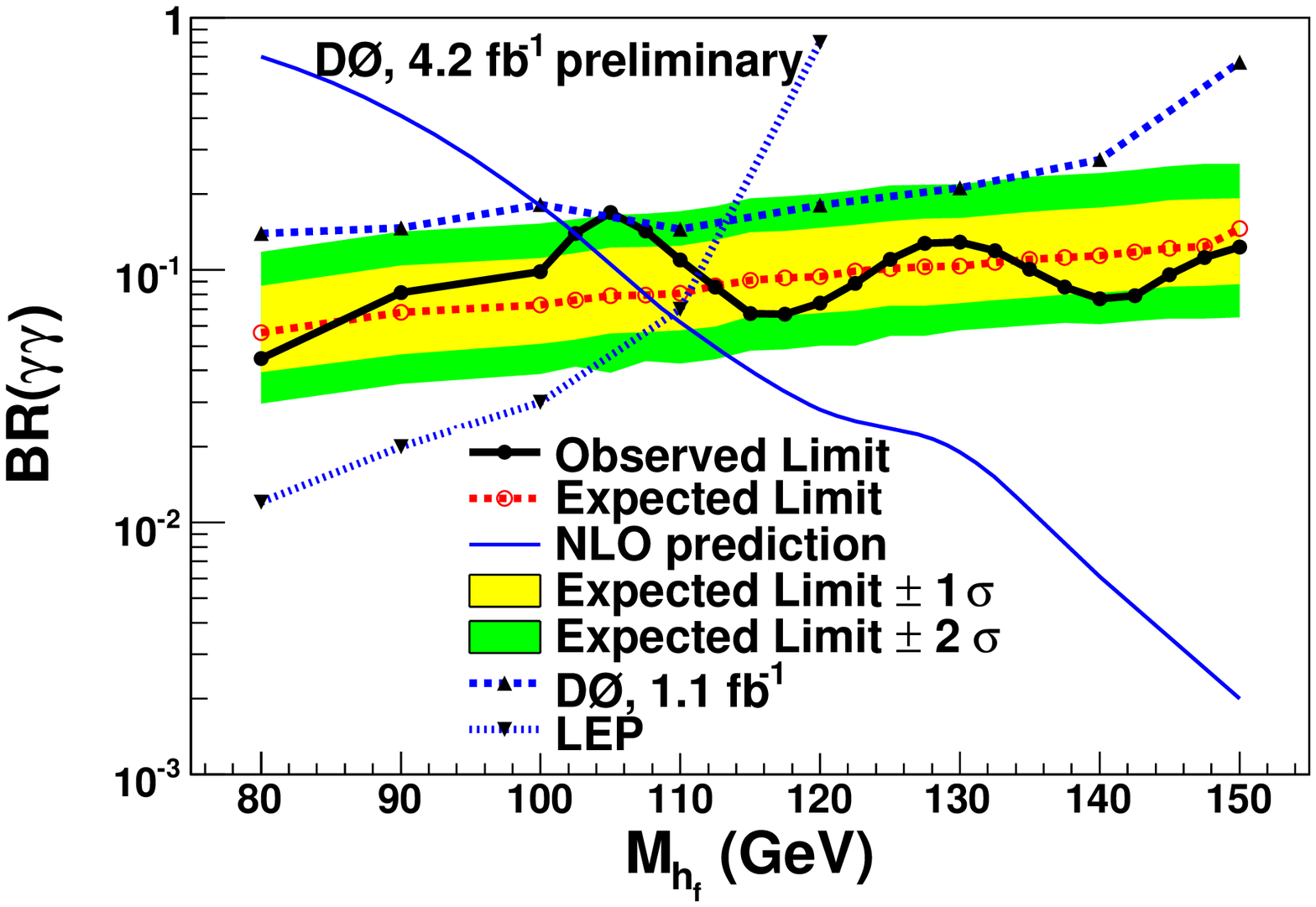}
&\includegraphics[height=.15\textheight]{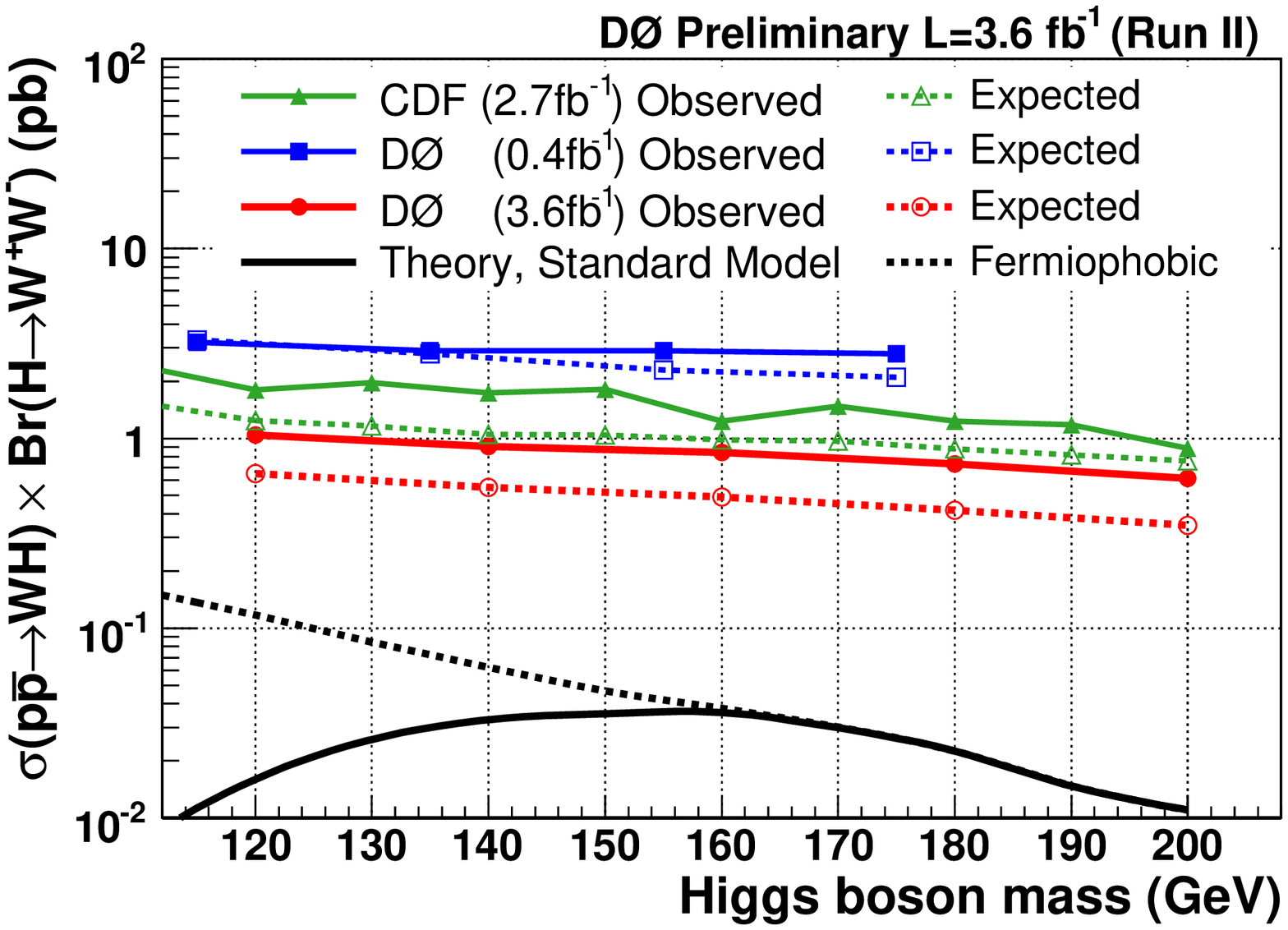}\\
(a) & (b)
\end{tabular}
  \caption{\label{fig:fermiophobic} 
(a) The limits on Br($\rm h_f\rightarrow\gamma\gamma$) as a function of
Higgs boson mass ($\rm M_{h_f}$); $\rm M_{h_f}<102.5$ is excluded at
95\% CL. (b) The 95\% CL limits on 
$\rm\sigma(p\bar{p}\rightarrow W^\pm h_f)\times Br(h_f\rightarrow W^+W^-)$
as a function of Higgs boson mass. The previous D\O\  results are also shown
for comparison.}
\end{figure}

In the inclusive $\rm h_f\rightarrow\gamma\gamma$ search, both 
both VH and VBF production modes are considered. The background events for
this channel are contributed by $\rm Z^0/\gamma^*\rightarrow e^+e^-$,
$\rm \gamma$+jet and QCD dijet production processes. Here the
invariant mass ($\rm M_{\gamma\gamma}$) distribution from two
reconstructed photons (with $\rm p_T>20$ GeV) has been utilized
as the final discriminating variable. The observed limits on the
production cross section have been translated into the benchmark model
expectations and a Fermiophobic Higgs boson mass below 102.5 GeV is
excluded at 95\% CL (see Fig.~\ref{fig:fermiophobic}(a)).

In the Fermiophobic Higgs models, the $\rm BR(h_f\rightarrow W^+W^-)$
becomes nearly 100\% for a Higgs boson mass above 100 GeV~\cite{hwwreference}.
The signatures of $\rm W^\pm h_f(\rightarrow W^+W^-)$ events
have been looked for in the events containing two like-signed
leptons, $l^\pm l^\pm$ (electrons or muons). The physics processes
like $\rm W^\pm Z^0\rightarrow$$l^\pm l^\pm l^\mp$ and
$\rm Z^0 Z^0\rightarrow$$l^\pm l^\mp l^\pm l^\mp$ along with the
misreconstructed $\rm Z^0/\gamma^*\rightarrow e^+e^-/\mu^+\mu^-$
events are the most dominant background processes for this search
channel. The observed and expected limits on
$\rm\sigma(p\bar{p}\rightarrow W^\pm h_f)\times Br(h_f\rightarrow W^+W^-)$
are compared to the SM and Fermiophobic model predictions in
Fig.~\ref{fig:fermiophobic}(b).
\section{Search for Neutral MSSM Higgs bosons}
The MSSM\cite{mssm1ref}\cite{mssmref2} requires two doublet
Higgs fields to generate masses to both ``up''-type
and ``down''-type fermions. After electroweak symmetry breaking,
such a two-Higgs-doublet model predicts the existence of five
physical Higgs bosons: two CP-even Higgs bosons, h and H, one
CP-odd Higgs boson, A and a pair of charged Higgs bosons, $\rm H^\pm$.
At the tree level, the MSSM Higgs phenomenology can fully be
described by two free parameters: $\rm m_A$, the mass of the 
CP-odd Higgs boson and $\rm tan\beta$, the ratio of vacuum 
expectation values of the Higgs fields to the ``up''-type
and ``down''-type fermions respectively.

The couplings of neutral Higgs bosons to bottom quark (b) and tau lepton
($\rm\tau$) i.e., ``down''-type fermions scale as $\rm tan\beta$  with
respect to their SM values. Therefore for high $\rm tan\beta$, the
production cross sections involving b quarks are enhanced of a factor
$\rm tan^2\beta$. Moreover, the CP-odd Higgs boson A becomes
degenerate with either one of the other neutral Higgs bosons, h/H which
leads to a total $\rm 2\cdot tan^2\beta$ enhancement in production
cross section relative to the SM prediction.
At high $\rm tan\beta$, the neutral Higgs
bosons ($\rm h/H/A\equiv\Phi$)
decay predominantly into $\rm b\bar{b}$ (Br$\approx$90\%)
or $\rm\tau^+\tau^-$ (Br$\approx$10\%) pairs. 

\begin{figure}[t]
\begin{tabular}{ccc}
\includegraphics[height=.15\textheight]{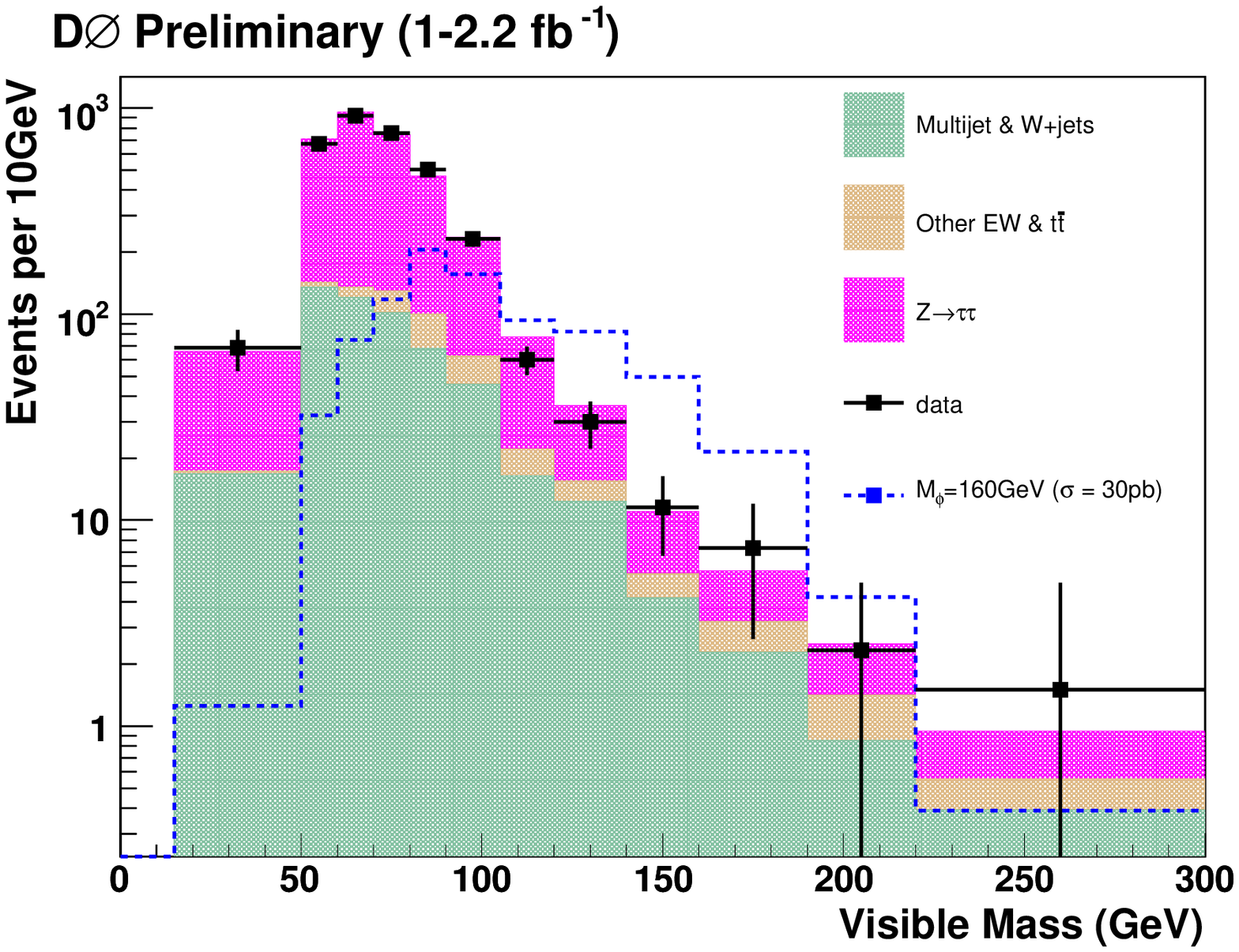}
&\includegraphics[height=.15\textheight]{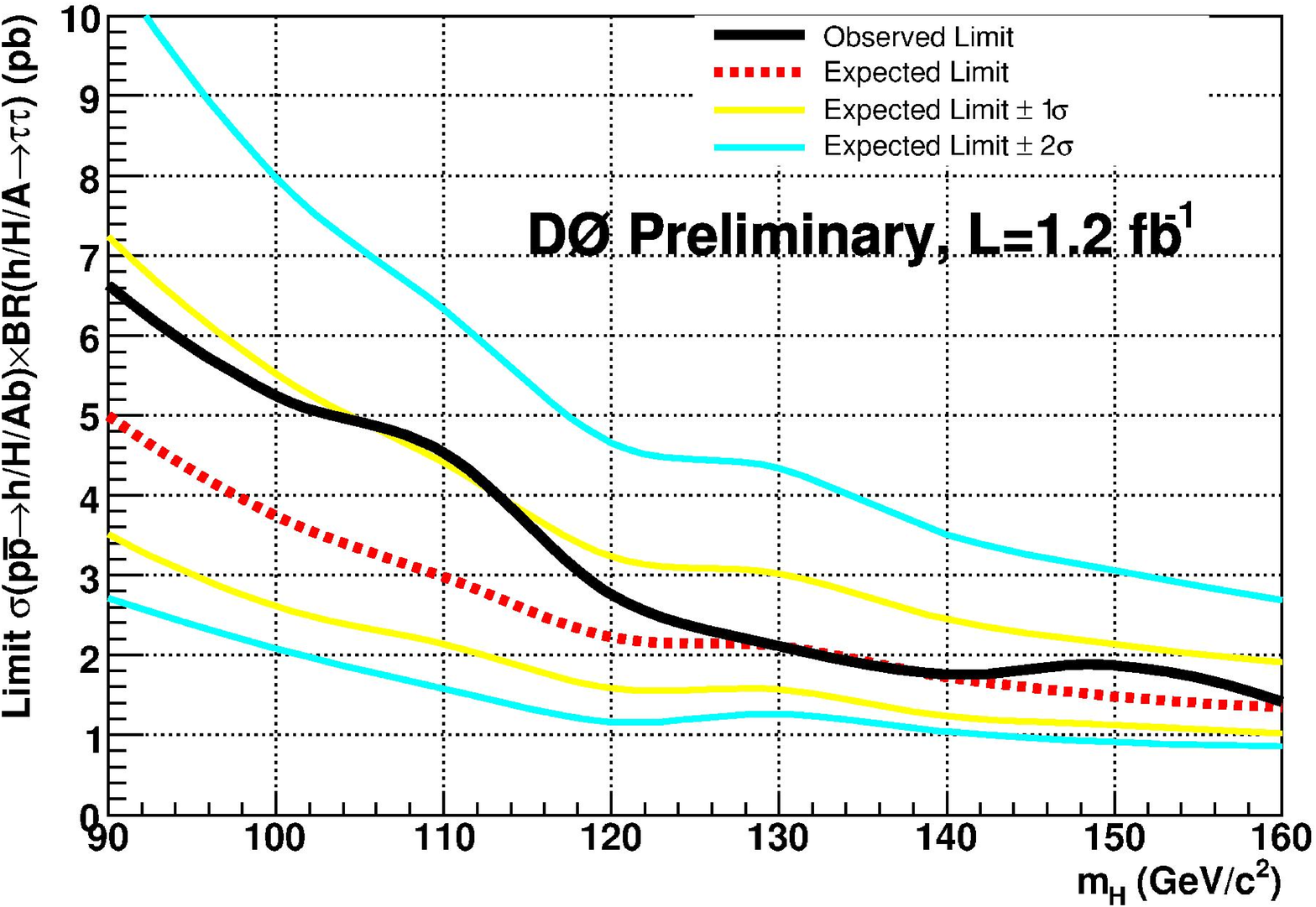}
&\includegraphics[height=.15\textheight]{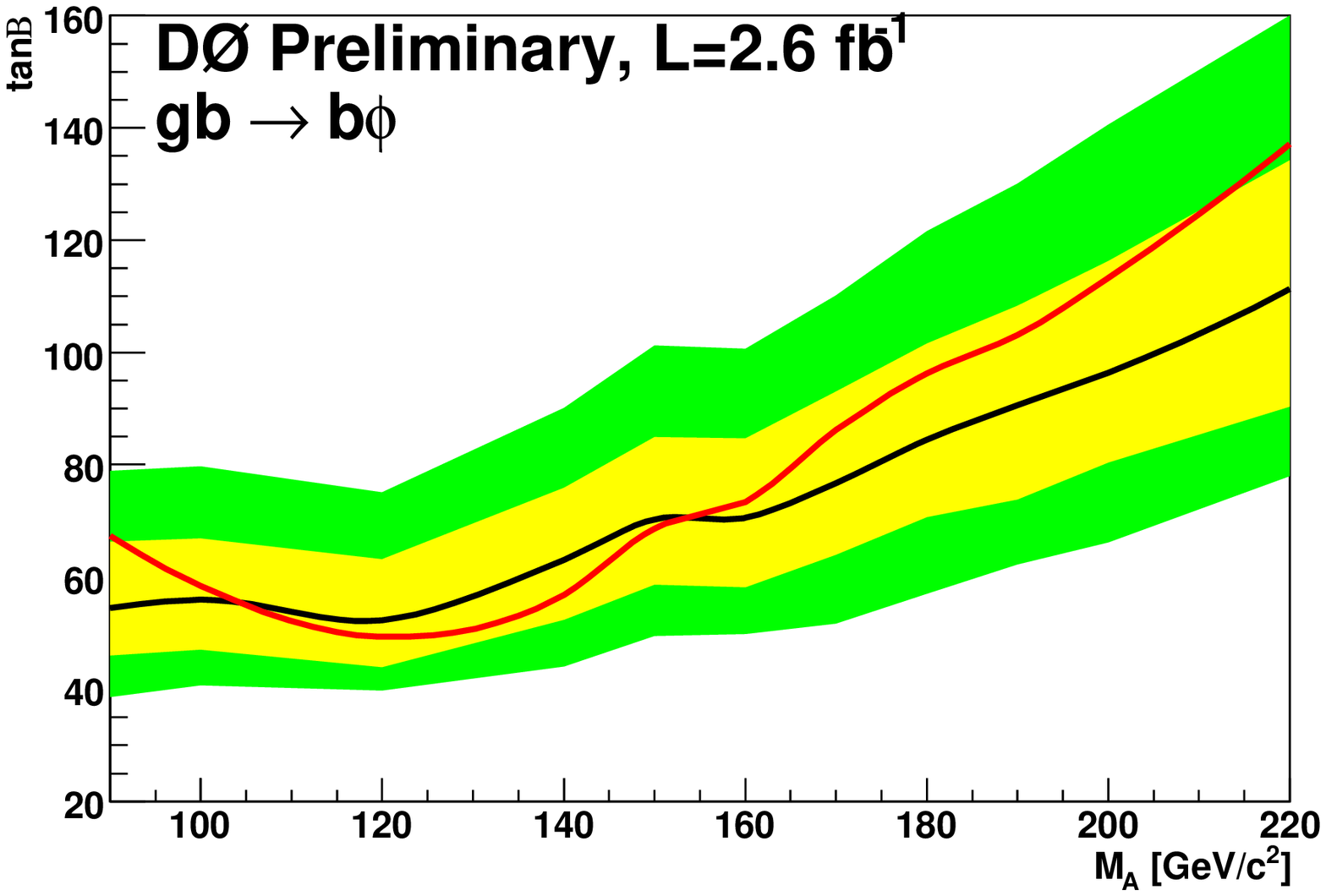}\\
(a) & (b) &(c)
\end{tabular}
  \caption{\label{fig:neutralmssm}
(a) The visible mass distribution (see text) for inclusive
$\rm p\bar{p}\rightarrow\Phi(\rightarrow \tau^+\tau^-)$ searches.
(b) Limits on $\rm \sigma(p\bar{p}\rightarrow\Phi b(\bar{b}))\times BR(\Phi \rightarrow \tau^+\tau^-)$
vs $\rm M_\Phi$. 
(c) The tree level limits on $\rm tan\beta-m_A$ from $\rm \Phi b(\bar{b})\rightarrow b\bar{b}b(\bar{b})$
search channel.}
\end{figure}

Three different analyses {\it viz.,} 
$\rm \Phi \rightarrow \tau^+\tau^-$, 
$\rm \Phi b(\bar{b})\rightarrow \tau^+\tau^- b(\bar{b})$
and $\rm \Phi b(\bar{b})\rightarrow b\bar{b}b(\bar{b})$
are presented here consisting of 2.2, 1.2 and 2.6 $\rm fb^{-1}$ datasets respectively.
The inclusive $\rm \Phi \rightarrow \tau^+\tau^-$ search
considers the events where one of the tau leptons decays
into a muon ($\rm\tau_\mu$) while the other tau lepton decays
hadronically ($\rm\tau_{had}$). The hadronically decaying
tau leptons are classified into three different types depending
on their detector signatures and are selected through the
usage of neural network discriminants ($\rm NN_\tau$). The
background events for this channel are dominated by
$\rm Z^0/\gamma^*\rightarrow\tau^+\tau^-$, $\rm Z^0/\gamma^*\rightarrow\mu^+\mu^-$ 
and multijet QCD production processes. The $\rm W^\pm$+jet events are
rejected by requiring $\rm M_T(p_T^\mu, \met)<40$ GeV.
The said analysis with 1.2 $\rm fb^{-1}$ Run IIb dataset is combined
with earlier analysis~\cite{runIIatautau} with 1 $\rm fb^{-1}$
of Run IIa dataset which considers additional final states {\it viz.,}
$\rm \tau_e\tau_{had}$ and $\rm \tau_e\tau_{\mu}$.
The distribution for the combined visible mass i.e., $\rm\sqrt{(P_{\tau 1}+P_{\tau 2}+\pet)^2}$,
where $\rm P_{\tau 1,2}$ is the four vector of the
visible tau decay products and $\rm \pet=(\met ,\metx , \mety ,0) $,
is shown in Fig.~\ref{fig:neutralmssm}(a).
In the $\rm \Phi b(\bar{b})\rightarrow \tau^+\tau^- b(\bar{b})$
search,
the events are characterized by an isolated muon, a hadronic
tau jet candidate (same $\rm NN_\tau$ selection as in 
$\rm \Phi \rightarrow \tau^+\tau^-$) and a b-tagged jet. 
Multijet QCD, $\rm t\bar{t}$ and 
$\rm Z^0/\gamma^*\rightarrow \tau^+\tau^- b/c$ are the dominating
background processes and they are reduced further by applying a
2D neural network discriminant. Fig.~\ref{fig:neutralmssm}(b) shows
the limits on  
$\rm \sigma(p\bar{p}\rightarrow\Phi b(\bar{b}))\times BR(\Phi \rightarrow \tau^+\tau^-)$
at 95\% CL.
In the $\rm \Phi b(\bar{b})\rightarrow b\bar{b}b(\bar{b})$ search,
the events are required to have 3-5 jets ($\rm p_T>20$ GeV and
$\rm |\eta|<2.5$) three of which are b-tagged. The overwhelmingly
large QCD background contributions are determined through usage of
both data and Monte Carlo (MC) distributions. The invariant mass
distribution from the jet pairs are considered here after applying
a likelihood discriminant for further separation between signal and
background processes. The cross section limits are translated into the
$\rm tan\beta-m_A$ space assuming a simple $\rm 2\cdot tan^2\beta$
enhancement over the SM and are shown in Fig~\ref{fig:neutralmssm}(c).
\begin{figure}[hb]
\includegraphics[height=.15\textheight]{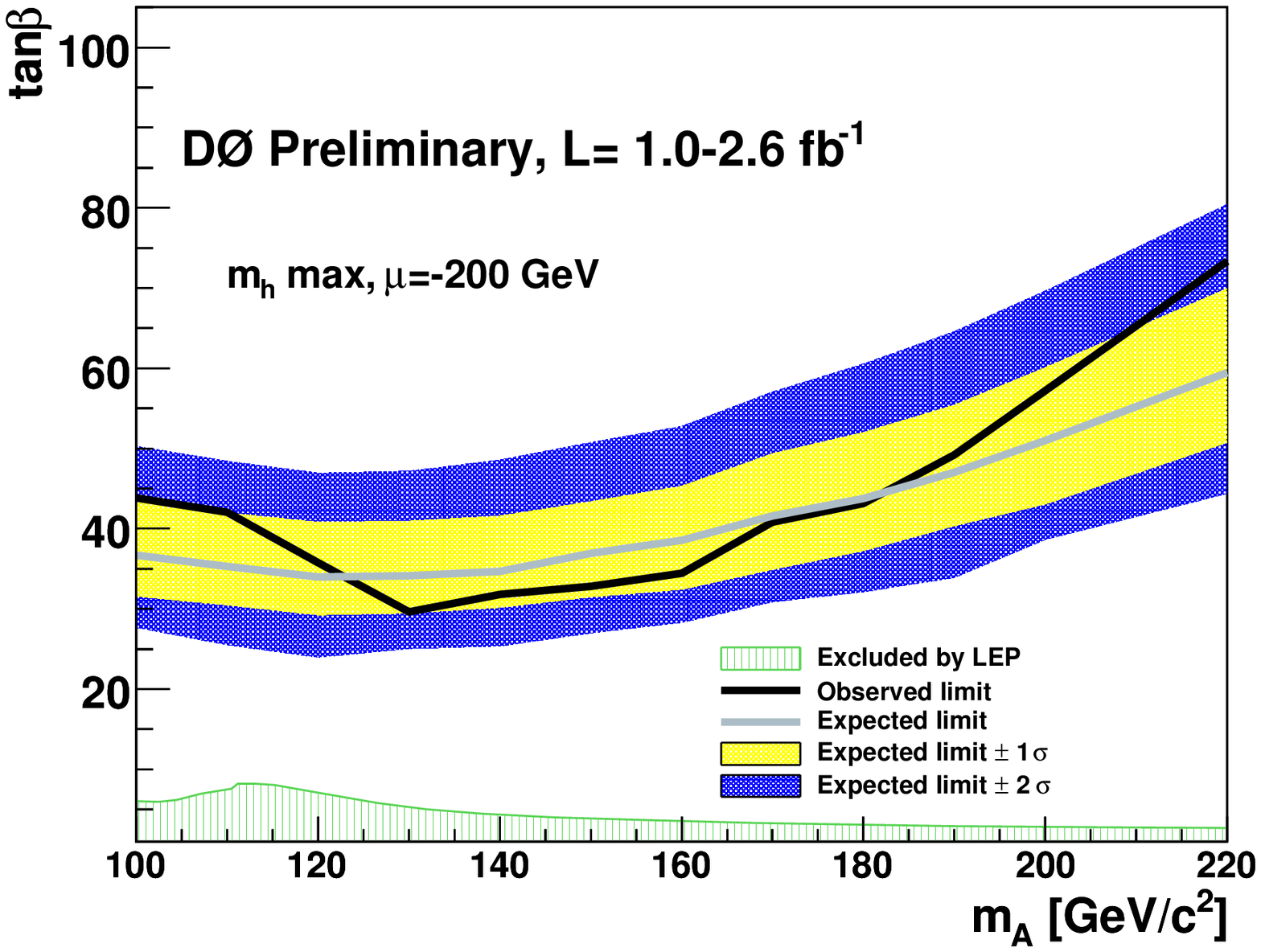}
\includegraphics[height=.15\textheight]{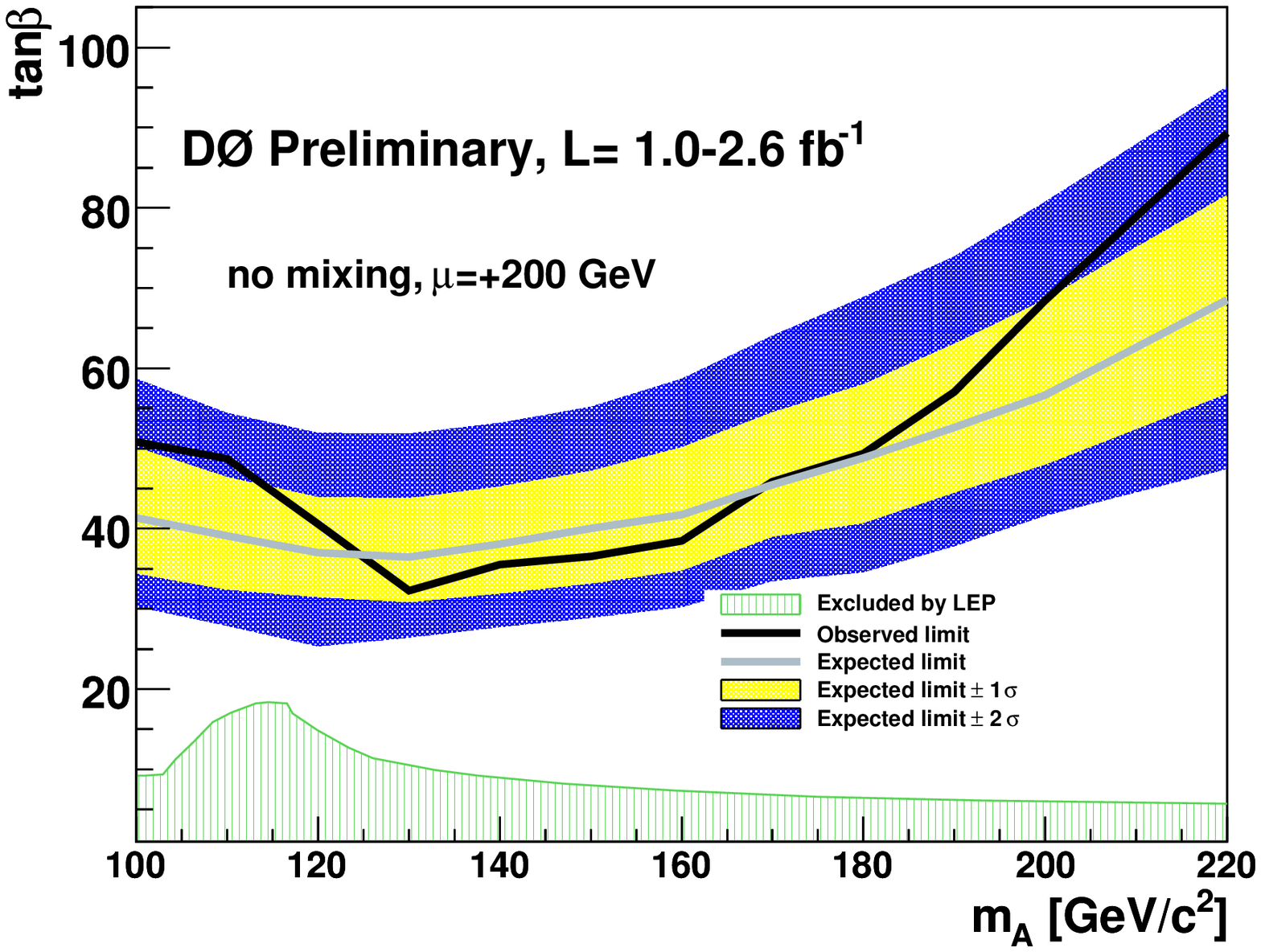}
  \caption{\label{fig:neutralmssmcomb}
Combined limits on MSSM parameter space for different benchmark scenario:
\hspace*{0.5cm}
$\rm m_h^{max},\mu$=-200 GeV (left) and no-mixing, $\mu$=+200 GeV(right).}
\end{figure}

Finally, all three above mentioned searches are combined taking into
account systematic uncertainties, including the correlations across
channels and analyses.
Fig.~\ref{fig:neutralmssmcomb} shows the combined
results interpreted in different benchmark scenarios\cite{mssmbenchmarkref}.

\section{NMSSM Higgs Search}
In the next-to-MSSM (NMSSM)~\cite{nmssmref}, at lower
masses the Higgs boson predominantly decays into a
pair of lighter neutral pseudoscalars, a.
The result is a suppression of $\rm h\rightarrow b\bar{b}$
branching ratio. Depending on its mass, the neutral pseudoscalar
can decay into a pair of muons ($\rm 2M_\mu<M_a<3M_\pi$)
or taus ($\rm 2M_\tau<M_a<2M_b$). Thus the final state signatures
would consist of two muon pairs or two tau pairs. However,
because of the difficulties with the 4$\tau$ final state, the
search is performed in the  
$\rm h\rightarrow a(\rightarrow\tau^+\tau^-)a(\rightarrow \mu^+\mu^-)$
final state.
Due to the extreme collinearity of the $\rm \mu^+\mu^-$ pair from
the pseudoscalar decay and the finite angular resolution of the D\O\ 
muon system, each muon pair in 4 muon channel is required to
contain one muon trajectory associated with a companion track.
A requirement of $\rm\Delta R>1$ between the muon pairs is applied to ensure
that these muon pairs are well separated from each other. For the
$\rm 2\tau+2\mu$ channel however both the muons in a muon pair are
required to be reconstructed and should be close to each other
within $\rm\Delta R<1$. It is to be noted here that $\rm 2\tau+2\mu$ 
analysis does not require explicit tau reconstruction (compromised
by the collinearity of the taus). Instead, the $\met$ requirements
are used to predominantly select the events with tau candidates.

For $\rm M_a < 2M_\tau$, the 4 muon channel sets an upper limit
of 10 fb on $\rm\sigma(p\bar{p}\rightarrow hX)×Br(h\rightarrow aa)×Br(a\rightarrow \mu^+\mu^-)^2$.
Assuming $\rm Br(h\rightarrow aa)\approx$100\% and the
SM $\rm\sigma(p\bar{p}\rightarrow hX)\approx$ 1.9 pb (for
$\rm M_h = 100$ GeV), the corresponding upper limit on 
$\rm Br(a\rightarrow \mu^+\mu^-)\approx$ 7\%.
However, in the NMSSM the $\rm BR(a\rightarrow\mu^+\mu^-)$ is expected to be larger
than 10\% for $\rm M_a<2m_c$ and depending on the branching ratio
of $\rm a\rightarrow c\bar{c}$ possibly even up to $\rm 2m_\tau$.
Thus the recently reported\cite{nmssmd0ref} D\O\  result severely
constrains the region $\rm 2M_\mu<M_a<2M_\tau$. For $\rm M_a>2m_\tau$,
the limits  set by the $\rm 2\tau+2\mu$
channel are a factor of 1-4 larger than the expected production
cross section.

\section{Conclusions}
D\O\  has made rigorous BSM Higgs bosons searches in
most of the promising channels. Neither of the searches presented
here observe a statistically significant excess over the SM
background expectations. Accordingly at 95\% CL, the most restrictive
limits have been established on relevant parameters for different theories.
Improved analysis techniques along with the rapidly increasing
datasets are expected to enhance the sensitivity of future BSM Higgs
searches. For the most recent results see Ref.~\cite{dzerorefpage}.

\bibliographystyle{aipproc}   

\end{document}